\documentclass[Journal]{IEEEtran}

\usepackage{lettrine}
\usepackage{amsmath}
\usepackage{systeme}
\usepackage{algorithm}
\usepackage[noend]{algpseudocode}
\makeatletter
\def\BState{\State\hskip-\ALG@thistlm}
\makeatother
\usepackage{algorithm}
\usepackage{lipsum, babel}
\usepackage{multicol}
\usepackage{stackengine}

\usepackage{extarrows}

\usepackage{textgreek}
\usepackage{blindtext}
\usepackage{graphicx}
\usepackage{color}
\usepackage{gensymb}
\usepackage{array}
\newcolumntype{P}[1]{>{\centering\arraybackslash}p{#1}}
\usepackage{tikz}

\ifCLASSINFOpdf
\else
\fi
\begin{document} 



\title{Secure and Privacy-Preserving Stored Surveillance Video Sharing atop Permissioned Blockchain}

\author{
\IEEEauthorblockN{Alem Fitwi, Yu Chen}
\IEEEauthorblockA{\\Dept. of Electrical and Computer Engineering,
Binghamton University, Binghamton, NY 13902, USA \\
Emails: \{afitwi1, ychen\}@binghamton.edu}
}

\maketitle
\begin{abstract} 
At present, more than a billion closed circuit television (CCTV) cameras are watching the world. These cameras garner a lot of visual information that is often processed and stored in remote and centralized cloud servers. Multiple occasions have revealed that this traditional approach is plagued with security and privacy breaches. The breaches could be interception of raw videos while in transit to distant surveillance analytics centers (SAC), infiltration to cameras and network video records (NVR), or abuse of cameras and stored videos. Hence, the traditional video surveillance system (VSS) cannot guarantee the protection of the privacy of individuals caught on CCTV cameras. Wherefore, this paper proposes a Secure and Privacy-preserving Stored surveillance videos sharing (SePriS) mechanism for authorized users/nodes based on smart contracts, blockchain (BC) and the enciphering of video frames using DAB, a mechanism developed based on discrete cosine transform (DCT), advanced encryption standard (AES), and block shuffling (BS) algorithm. The BC-based solution creates an environment auspicious for creating a decentralized, reliable SACs and storage sites  with secure and privacy-aware sharing of stored surveillance videos across SAC nodes and by law enforcers, police departments, and courts securely connected to the SAC nodes. The experiments and analyses validate that the proposed BC-based SePriS solution achieves the design purpose. 

\end{abstract}
\IEEEpeerreviewmaketitle
\section{Introduction}
\label{sec:intro}

To ensure public safety and physical security, the number of closed circuit television (CCTV) cameras deployed in urban and suburban areas have been increasing quickly over the past decade. As projected by many experts and indicated in recent reports \cite{lin2019a}, more than a billion of CCTV cameras are watching the world today. Therefore, these cameras enable the law enforcers and security personnel to collect tremendous amount of information about individuals without their knowledge and consent while moving along many public and private places. The pros of video surveillance systems (VSS) might outweigh the cons; however, there are two valid arguments raised by two antagonistic groups. On one hand, there are people who advocate the widespread deployment and use of VSS because it helps increase public safety and home security, and reduce crime rates. On the other hand, there are a number of people who argue against the practice of VSS in public places because they strongly believe that it could be abused so easily. 
CCTV cameras have become commonplace around the world where individuals are observed numerous times a day without their consent and knowledge. There are a myriad of reports that substantiate the fear of abuse of CCTV cameras. The German Chancellor, Angela Merkel, was spied while at her apartment by abusing a museum security CCTV camera \cite{cavallaro2007privacy, fitwi2019agent, fitwi2020prise}. Quite recently, on March 9, 2021, a group of hackers claimed to have hacked the security CCTV camera data from Verkada and gained access to the live feeds as well as central storage of 150,000 surveillance CCTV cameras deployed in a number of companies and hospitals, including Halifax Health \cite{Jackie2021}. 

In general, privacy breaches are mainly attributed to the interception of raw video streams that contain visual information about individuals by adversaries/intruders and abuse of videos by the people in charge of the surveillance. The abuse of the surveillance system include collecting unauthorized data about individuals by maneuvering the cameras, accessing and leaking collected data or threatening to do so, and blackmailing individuals caught on cameras \cite{rajpoot2015video, wang2018enabling}. Our previous works focused on privacy-sensitive attributes detection and ensuring end-to-end (E2E) privacy \cite{fitwi2020privacy, fitwi2021privacy, fitwi2020minor}. However, all of the aforementioned problems cannot be addressed by introducing good privacy-attributes detection and scrambling schemes alone. The addition of effective access management is vital to ensure privacy. Stored videos must be accessed with clear authentication, authorization, and accounting policies in place. 

A permissioned, private blockchain (PBC) network is envisioned in this work to alleviate the privacy and security problems vis-à-vis access to surveillance videos processed and stored in distributed surveillance analytics centers (SAC) and storage sites, respectively. It employs privacy-preserving video-frame encrypting mechanisms and smart contracts that define privileges and access rules, which allow only authorized users to access the videos without violating the privacy of individuals. The existing smart contracts, however, do not preserve the privacy and confidentiality of data when shared to the BC nodes. In our proposal, the smart contract is tailored for better handling of the privacy and confidentiality issues. Users who try to access the stored videos need to prove themselves to the blockchain that they have the required privileges. Besides, correctness of the references are validated by every other node so that only authorized accesses are made and the integrity of the video storage is verifiable. The reference in this paper points to a variable on a mapping table that contains actual references to the videos. On top of that, the video frames are exchanged in scrambled form to prevent spilling of personal information by interception attacks. 

To prevent the abuse of stored surveillance videos, we proposed and experimentally validated a Secure and Privacy-preserving Stored surveillance videos sharing (SePriS) mechanism, a PBC-based solution along with an efficient video-frame scrambling mechanism. Our major contributions are briefly outlined as follows:
\begin{itemize}
		\item Design of PBC-based video surveillance system that ensures secure and private exchange of stored video frames to prevent leaking and blackmailing of individuals caught on CCTV cameras. SePriS ensures that videos are accessed only in an authenticated, authorized, and accountable way. 
		
		\item A video-frame enciphering mechanism named DAB in introduced based on discrete cosine transforms (DCT), advanced data encryption standard (AES), and a block shuffling (BS) algorithm to ensure secure exchange of video streams off-BC storage sites and users. 
		
		\item Extensive experimental study and security analyses substantiate the validity and applicability of the proposed  SePriS architecture, mechanisms, and techniques.
		
\end{itemize}
The remainder of this paper is organized as ensues: the related works are introduced in Section \ref{sec:relw}. The overall system architecture of SePriS is then presented in Section \ref{sec:sarch}. In Section \ref{sec:pbc}, the blockchain based solutions are formulated and described followed by the description of DAB, the DCT-AES-BS based video-frame enciphering scheme in Section \ref{sec:enci}. The analysis and discussion of the proposed SePriS scheme are presented in Section \ref{sec:rad}. Eventually, the conclusions are presented in Section \ref{sec:con}.
\section{Related Works}
\label{sec:relw}
\subsection{Video Surveillance Systems and Privacy}

Today, more than a billion of CCTV cameras are pervasively deployed around the world in urban areas by governments, companies, and individuals with the main goals of ensuring physical security and public safety. In places like Beijing, London, and New York, an individual is estimated to be caught on CCTV cameras hundreds of times a day \cite{fitwi2019agent, fitwi2020minor, wang2018enabling, yuan2020minor}. Consequently, a tremendous amount of information about individuals is garnered without their knowledge and consent. This huge data containing privacy-sensitive visual information could be divulged into the wider cyber space where there are almost 4.57 billion active Internet users as of July 2020 that accounts 59\% of the global population \cite{clement2020global}. The spill of the information is often attributed to interception attacks and abuse of cameras and stored videos for blackmailing, cyber stalking, or extorting. This poses a great risk of breaches and invasions of the privacy of individuals caught on CCTV cameras. The best way to create a video surveillance system that protects the privacy of individuals amid the increased proliferation of CCTV cameras is requiring the camera manufactures to incorporate privacy-preserving methods by design and recommend secure deployment networks \cite{altawy2017security, cavoukian2012privacy, vattapparamban2016drones}. 

Many incidences have been reported in relation to privacy breaches and CCTV camera approaches. About a quarter of century ago, in August 1995, numerous peeping incidences or voyeurism to observe or record people doing private stuffs at their homes were reported \cite{goldberg1995introduction}. Unfortunately, those bad practices and abuses did not stop there; rather they have continued in a more sophisticated way. Even government officials' apartments were spied by abusing and directing CCTV cameras via windows \cite{cavallaro2007privacy}. The American Civil Liberties Union (ACLU) have identified and reported multifaceted CCTV camera abuses including criminal abuse, institutional abuse, abuse for personal purposes, discriminatory targeting, and voyeurism \cite{senior2005enabling}. 

Generally, the public has been repeatedly informed that surveillance cameras are never abused by authorized people in charge of them. Hence, many a person might believe that people who sit in the surveillance operation centers (SOC) can supposedly be trusted not to abuse these CCTV cameras at their disposal, which boost their sighting capability drastically, to engage in despicable or outright illegal behavior. Nonetheless, this understanding has been proven to be glaringly false as the report of literally multiple incidences clearly indicate \cite{Notbored2010}. A number of protests against the use of CCTV cameras were conducted in the best, where a list of many such protests against surveillance cameras are provided in reports citing the abuses and the ineffectiveness of video surveillance as a ``crime-fighting'' tool \cite{Notbored2009}. Even today, CCTV cameras are not secure and they greatly risk the privacy of many individuals.
\subsection{Blockchain Technology}

About a decade ago, the Bitcoin has substantiated how the BC technology can enable decentralize trusted computing models \cite{nakamoto2008bitcoin}. The BC technology as a whole was made popular by the success of Bitcoin. Now it can be employed to make trustworthy and secure transactions across untrusted networks without relying on a trusted centralized third party. The BC is a kind of chronological sequence of blocks including a list of complete and valid transaction records. The blocks that constitute the BC blocks are chained to one another by a hash value, where the block preceding a given block is called its parent block, and the first block is known as the genesis block. Each block comprises a block header and a block body. The block header contains the block version, previous block hash, timestamp, 4-byte nonce, body root hash, and target hash. The block body consists of validated transactions within a specific time period. These features enable the BC technology to have the potential to decentralize many applications that depend on a centralized trusted body. 

Moreover, a light-weight blockchain along with the concept of identity-based distributed data possession in multi-cloud storage can be leveraged to ensure privacy and authorized access in many smart applications; additional works are required, though \cite{kshetri2017blockchain, xu2019microchain,  xu2020decentralized}. However, due to concerns on privacy and performance issues, a public blockchain is not an ideal candidate. Instead, a private blockchain has been considered where only authenticated member nodes join \cite{do2017blockchain, zyskind2015decentralizing}. Therefore, a lightweight, closed-group blockchain that supports decentralized applications like surveillance that entails high speed and privacy could meet the requirements \cite{xu2020blendsps}. The main issue is speed. Video surveillance is a real-time process but the BC doesn't allow this yet. As a result, the BC technology is employed in SePriS mechanism to manage accesses to videos stored in many distributed sites. 

\subsection{Video Frame Scrambling Mechanisms}

There are a multitude of schemes like \textit{editing, face regions, false color, and JPEG} \cite{rakhmawati2018image} that can be employed to scramble or mask video frames. However, scrambling/encryption scheme is the only feasible option in the eyes of security/privacy and usability. It securely hides sensitive information on video frames and prevents unauthorized accesses. All other schemes are not able to meet the stringent requirements of video-contents privacy protection as they fail to achieve an optimal balance between privacy, clarity, reversibility, security, and robustness. 

Most of the preexisting encryption mechanisms are not suitable for enciphering video frames due to their bulky nature. Public-key cryptographic schemes like Rivest-Shamir-Adleman (RSA) and Elliptical Curve Cryptography (ECC) are too slow to be used for image scrambling due to their heavy reliance on hard factoring, a compute-intensive process. The conventional encryption mechanisms like Rivest Cipher 4 (RC4) and Advanced Data Encryption (AES) are not suitable for video encryption, either. AES is one of the most secure encryption scheme employed in the transport layer security (TLS) in the TCP/IP Network Architecture. It, nonetheless, has some issues with video/image scrambling like inability in breaking the strong correlation amongst adjust pixels. Chaotic schemes are said to be the best mechanisms for video scrambling due to their high randomness, sensitivity to slight change, and ability to be vectorized. Their shortcoming is the fact that they use XOR operator for mixing the chaos and plain image; hence, a separate key for every frame is required to meet perfect secrecy. In this paper, we proposed DAB, a video encryption mechanism that employs DCT, AES, and block shuffling to reduce the number of keys employed in chaotic schemes, and to improve the correlation problems observed in AES ciphers when used for image encryption.
\section{SePriS System Architecture}
\label{sec:sarch}
\begin{figure*}[t]
    \centering
        \includegraphics[width=0.85\textwidth]{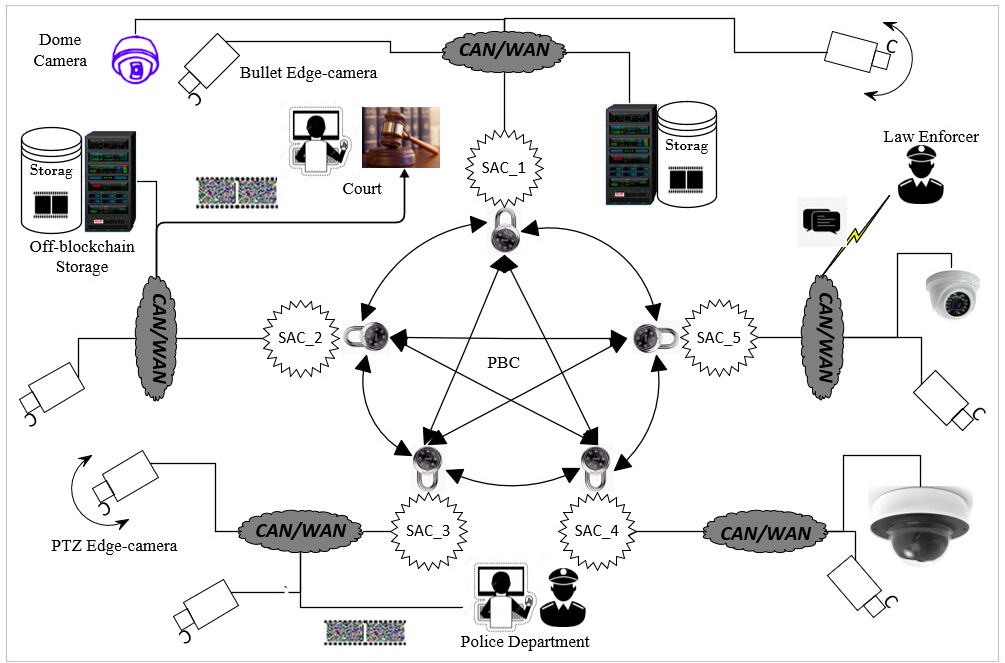}
    \caption{PBC-based SePriS Architecture for secure and privacy-aware exchange of surveillance videos stored in distributed off-BC sites.}
    \label{pbcarch}
    \vspace{-10pt}
\end{figure*}

It is feasible to use BC technology to address the problems in distributed access to stored surveillance videos. It enables the construction of a privacy-aware smart surveillance system by integrating the advanced features of BC and smart contracts that define how security, privacy, integrity, authentication, authorization, controllability, auditability, and accounting can be achieved. Our SePriS system comprise three major actors, namely off-BC distributed video storage sites, BC nodes connected to SACs, and users. The off-BC storage refers to secure cloud/fog based storage system for the storage of surveillance videos connected to the BC nodes. The BC network ensures authenticity. Users are assigned with different levels of access privileges to the videos, which are defined in the smart contracts. Users could be security personnel working in SOCs, law enforcers, police department, or courts.

Figure \ref{pbcarch} portrays the high-level overview of the proposed SePriS system, which comprises smart cameras as edge devices. Authorized users and off-BC storage connected to the BC-nodes through a local area network (LAN) or a wide area network (WAN). Fine-grained access control policies and privileges are set as part of the privacy policies. They can be legitimately and authorizedly updated or revoked at any time. References to access histories of stored or viewed videos are logged and posted onto the BC based on the details defined in smart contracts. Viewer's identity, access time, privileges, and references of accessed parts are also logged and stored. To discourage leaking of videos and images, information specific to the viewers, like fingerprint, is appended into the log-reference of every watched video and shared with all BC-nodes. The details are presented in Section \ref{sec:pbc}. 

Here two type of cryptographic schemes are employed to ensure secure and privacy-conscious exchange of messages and video frames. A digital signature is embedded in requests for access to stored videos or messages sent to any of the BC-nodes by any of the users For this purpose, an elliptic curve public encryption/decryption (ECPED) is employed. Following identity and security authentication, consensus by the BC-nodes, and the successful granting of access-code, access to videos is granted to the requestor on the off-BC storage site in scrambled form to ensure end-to-end (E2E) privacy. The scrambling is performed by DAB scheme described in Section \ref{sec:enci}. 

\section{Permissioned Blockchain based Secure Access}
\label{sec:pbc}
In SePriS system, a PBC-based solution is introduced for secure and privacy-aware sharing of stored surveillance videos across different SACs and authorized users. Each of the SAC node backups the BC copy synchronously. The data stored in the BC comprises only access requests, users information, access control list, references to accessed videos associated with the identity information of the user who was granted access, and other activities performed on the video. The actual video frames are stored on distributed off-BC storage clouds as depicted in Fig. \ref{pbcarch}.

\begin{figure}[ht!]
    \centering
        \includegraphics[width=0.45\textwidth]{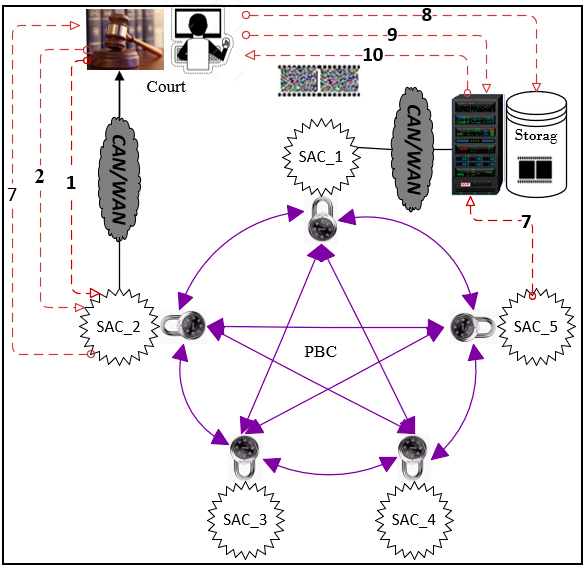}
    \caption{The work-flow of the PBC-based Solution for secure and privacy-aware exchange of stored surveillance videos.}
    \label{pbccon}
    \vspace{-10pt}
\end{figure}

Figure \ref{pbccon} illustrates how stored surveillance videos are securely accessed by authorized users using the PBC-based approach. Primarily, the police department, courts, and other law enforcers are the users of stored surveillance videos for investigative and forensic purposes, or as evidence against criminals in courts of law. Here, in the interest of clear demonstration, we will consider how a court can access stored videos securely from an off-BC storage site. As indicated in Fig. \ref{pbccon}, to access a video of specified date and range from an off-BC storage cloud, it takes 10 steps executed by the requestor, BC nodes, and the off-BC storage. Each of the steps are briefly described in what ensues:

\noindent\textbf{Step-1:} The court first identifies itself to any of the BC-nodes ($SAC\_2$ in this case) as portrayed in Fig. \ref{pbccon}. That is, an identity and a security authentications are first performed. The identity authentication involves bio-metrics where the requestor scans any of their fingers whose processed copies are already in the BC-nodes. If the authentication process succeeds, the requestor receives a unique identifying ID (for instance UID = court322874352017640022980892363199962446587) locked in a double lock-box as shown by Eq. \ref{eq:step1}, where $E$ stands for an asymmetrical encryption, $key_{pub-R}$ stands for the public key of the receiver (the court in this case), and $key_{pri-S}$ stands for the private key of the sender ($SAC\_2$ in this case). A copy of the UID is saved in the BC-nodes for later verification. The assumption is that the sender and receiver has each other's public keys. Then, the requestor proceeds to step-2.
\begin{equation}   
    \label{eq:step1}
      \begin{split}
        cipher\_UID\ =\ E(key_{pub-R}, E(key_{pri-S},UID))
      \end{split}
\end{equation}

\begin{figure}[b]
    \vspace{-10pt}
    \centering
        \includegraphics[width=0.5\textwidth]{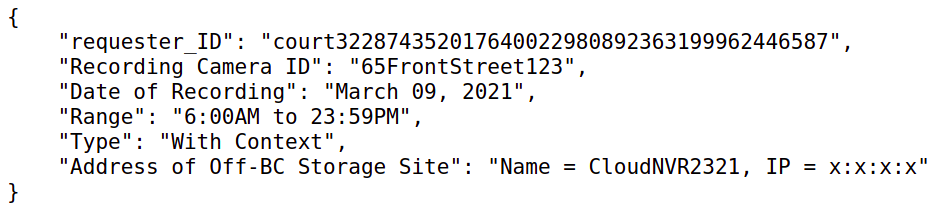}
    \caption{A sample request made by court, in JSON file format, sent to a BC-node.}
    \label{rjson}
\end{figure}

\noindent\textbf{Step-2:} A request for access to a specific video in the off-BC storage is forwarded to $SAC\_2$ enclosed inside double lock-boxes, described by Eq.\ref{eq:step1}. In other words, the sender initiates a transaction by sending a request signed by its private key and encrypted by receiving BC-node's public key. The request is easily verified by the receiving BC-node ($SAC\_2$) using the sender’s public key. Figure \ref{rjson} shows a sample request that comprises UID of a sender, ID(s) of the recording camera(s), Date of recording, range or time length of requested video, type that describes whether the sender wants the video with the whole context or specific behavioral activities identified using a machine learning or Deep learning model, and name and address of the off-BC storage site.

\noindent\textbf{Step-3:} The request is broadcasted by $SAC\_2$ to all other BC-nodes through the P2P network as shown in Fig. \ref{pbccon}.

\noindent\textbf{Step-4:} Each BC-node verifies the request in accordance with the predefined smart contract and database of legitimate users along with their access privileges, also called access control list (ACL). 

\noindent\textbf{Step-5:} Reach consensus and pack the validated request chronologically. That is, if the request is valid, it is appended to the chain as a new block following the solving of the proof of work (PoW) puzzle by a miner.

\noindent\textbf{Step-6:} Generate an access code for the requestor following consensus. $courtToCloudNVR232139869614605459$ is a sample access code generated for a court to access a video from the off-BC storage specified in the request.

\noindent\textbf{Step-7:} The generated $access code$ is forwarded to the court wrapped with the private and public keys of $SAC\-2$ and the court, respectively. Likewise, the original request in JSON file format is updated with the access code and sent to the target off-BC storage site in double boxes locked with the private and public keys of $SAC\_5$ and the off-BC storage, respectively.

\noindent\textbf{Step-8:} The requestor, court, performs an identity and a security authentication on the off-BC storage. 

\begin{figure}[t]
    \centering
        \includegraphics[width=0.48\textwidth]{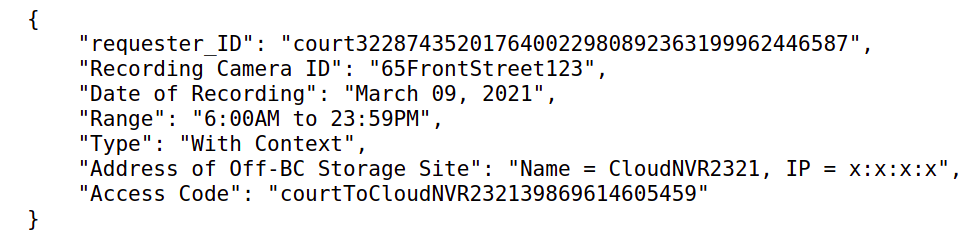}
    \caption{A sample request made by court, in JSON file format, an off-BC storage.}
    \label{acode}
    \vspace{-10pt}
\end{figure}

\noindent\textbf{Step-9:} Following a successful authentication, the requestor sends an updated request, shown in Fig. \ref{acode}, which contains the access code provided by the BC members in addition to the contents of the original request sent to the nodes in Step-2. Just like the previous cases, this request as well is encrypted and signed by the private and public keys of the court agent and the off-BC storage, respectively.

\noindent\textbf{Step-10:} If the request sent by the court in Step-9 and the request sent to the off-BC storage by the BC-nodes match, access to the specified video is granted as illustrated in Fig. \ref{pbccon}. Unlike the requests and short messages which are enciphered by using computationally expensive asymmetrical encryption mechanism, the video data cannot be encrypted using ECC for it contains bulky information. As a result, the video streams are encrypted using a DCT-AES-BS mechanism designed for this purpose and described in Section \ref{sec:enci}. In parallel, log of activities and references to videos accessed associated with requestor identity and device information are sent to the BC. They are permanently and synchronously stored in the BC. 

\section{DAB: DCT-AES-BS Based Video Frame Enciphering}
\label{sec:enci}
\begin{figure}[ht!]
    \centering
        \includegraphics[width=0.45\textwidth]{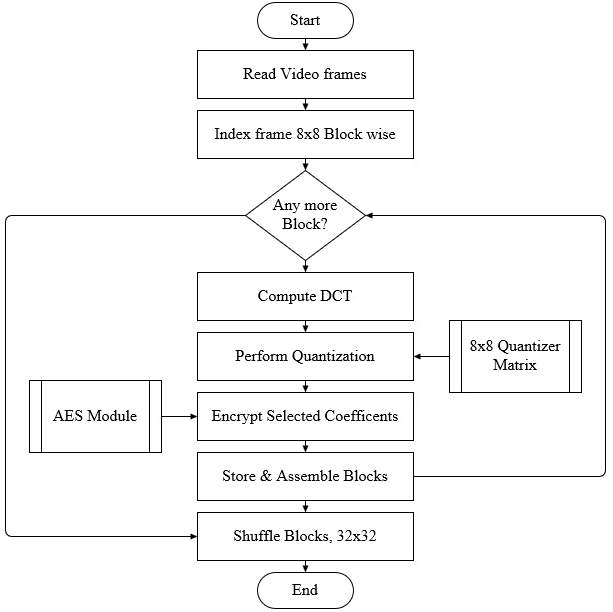}
    \caption{Video Frame Enciphering Process Flows of DAB.}
    \label{dctaes}
    \vspace{-10pt}
\end{figure}

Asymmetrical Encryption schemes like ECC and Rivest–Shamir–Adleman (RSA) are too slow to be employed for video encryption which contains bulky information. Rather they are used for enciphering short messages and creating digital signatures. As a result, DAB, a mechanism convenient for video enciphering is purposed in this section. It's meant for encrypting video frames accessed by authorized users from any of the off-BC storage sites shown in Fig. \ref{pbcarch}. Figure \ref{dctaes} presents the proposed DAB scheme that comprises four major modules, namely discrete cosine transom (DCT) calculator, quantizer, advanced encryption standard (AES), and a block shuffler (BS). They are briefly described in the following four subsections.

\subsection{DCT}
DCT is a well-known mathematical transformation technique that takes a signal and transforms it from spatial domain into frequency domain. Today, a number of digital image and video compression techniques employ a block-based DCT due to the fact that this technique reduces the amount of data needed to recreate a digitized image. In particular, JPEG and MPEG use the DCT to concentrate image information by removing spatial data redundancies in two-dimensional images \cite{coconu2002distributed}. In the process of the DCT, the input video frame is decomposed into $8\times 8$ blocks and the DCT of every $8\times 8$ block of video frame is computed by using Eq. \ref{eq:dct}. Technically speaking, these blocks are transformed from the spatial domain to the frequency domain representation by the DCT. 
\begin{equation}   
    \label{eq:dct}
      \begin{split}
        DCT\ &=\ T\times M\times T'\\
      \end{split}
\end{equation}
\noindent where $M$ is an $8\times 8$ block from an input frame, and $T$ is an $8\times 8$ DCT matrix computed by using the following equation,  
\begin{equation}
  T_{i, j}=\left\{
    \begin{array}{ll}
      \frac{1}{N}, & \mbox{if $i = 0$}.\\
      \sqrt{\frac{2}{N}}cos[\frac{(2j+1)i\pi}{2N}], & \mbox{if $i > 0$}.
    \end{array}
  \right.
\end{equation}

\subsection{Quantization}
\begin{figure}[t]
    \centering
        \includegraphics[width=0.4\textwidth]{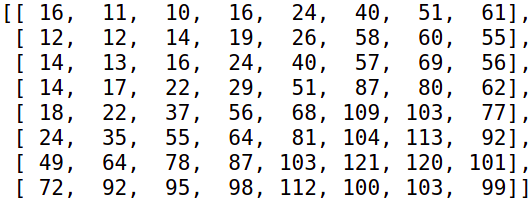}
    \caption{IJG Base Quantization Matrix.}
    \label{bqm}
    \vspace{-10pt}
\end{figure}

The $8\times 8$ DCT coefficients are now ready for quantization. Each DCT coefficient is divided by its corresponding constant in a standard quantization matrix, where values are rounded down to the nearest integers. There is currently one standard employed for computing the quantization matrix that has been around for a while ever since JPEG was proposed by the Independent JPEG Group (IJG). It depends on the Quality factor (Q). The basic algorithm is as follows: first a quality factor Q, which can assume a value from 1 to 100, is selected. 1 is a value of the ``poorest'' quality while 100 is the value of the ``highest'' quality. 50 is the default setting. The base IJG quantization matrix (bqm) based on which a desired quantization matrix is derived is provided in Fig. \ref{bqm}. 

A quantization matrix with a certain level of quality, Q, is computed by using Algorithm \ref{qma}. A parameter $S$ is computed based on whether $Q < 50$ or $Q>=50$ and is used as multiplier to the base quantization matrix to obtain the target quantization matrix.

\begin{algorithm}
\caption{Quantization Matrix}\label{qma}
\begin{algorithmic}[1]
\State $\textit{Base Quantization Matrix, bqm} \gets \text{Fig. \ref{bqm}}$
\State $\textit{Q} \gets \text{Select values: 1 to 100}$
\If {$Q < 50$}
    \vspace{5 pt}
    \State $\textit{S = } \gets \text{5000/Q}$
    \vspace{5 pt}
\Else 
    \State $\textit{S = } \gets \text{200-2*Q}$
\EndIf
\State $qm =  \gets \frac{floor(S\times bqm +50)}{100}$
\end{algorithmic}
\end{algorithm}
            
\subsection{AES}
In each block, there are 64 DCT coefficients set up from the lowest frequency at the upper left corner to the highest frequencies at the lower right corner. The most important visual characteristics of the frame are placed in the low frequencies while the details are situated in the higher frequencies. The Human Visual System (HVS) is most sensitive to lower frequencies than to higher ones \cite{pennebaker1992jpeg}.

Therefore the AES, the most widely used symmetrical encryption scheme in today's Internet, is employed in this work to encipher selected coefficients of every DCT block. Only those coefficients on locations (indices) [0][0], [0][1], [0][2], [0][3], [1][0], [1][1], [1][2], [1][3], [2][0], [2][1], [2][2], [2][3], [3][0], [3][1], [3][2], and [3][3] with higher HVS sensitivity in every DCT block are encrypted using AES.

\subsection{Block Shuffler}

Once all values of the selected locations in every DCT block are encrypted using AES, the whole frame is shuffled by using a simple but secure shuffling algorithm purposed in this section. It further randomizes the outputs of AES in block of size $32 \times 32$. AES coupled with the shuffling algorithm described in Algorithm \ref{sfl}. Hence, two keys are employed here: one is AES encryption key, and the other is index\_key used to recover shuffled blocks. Both are securely shared with the recipient using the ECC public key cryptographic method.
\begin{algorithm}
\caption{Block Shuffler\label{sfl}}
\begin{algorithmic}[1]
\State $\text{bs} \gets \text{blk\_size}$
\State $\text{frame} \gets \text{vid.read(0)}$
\State $\text{W, H} \gets \text{frame.size}$
\Procedure{shuffle\_frame}{frame, bs, H, W}
    \State $\text{hw} \gets \text{x, y tuples of each 32xs32 block}$
    \State $hws \gets Yates-shuffle(hw)$
    \State $index\_key \gets hws.index()$ 
    \State $hws \gets array(hws)$
    \State $ hws \gets hws.reshape(int(H/bs),\ int(W/bs))$
    \State $ imsize \gets frame.shape$
    \State $ imgn \gets frame$
    \State $ c1\gets 0$
    \For{\texttt{i in r\_[:imsize[0]:bs]}}
        \State \texttt{c2=0}
        \For{\texttt{j in r\_[:imsize[0]:bs]}}
        \State \texttt{x, y = hws[c1][c2]}
        \State \texttt{imgn[i:(i + bs),j:(j+ bs)] =}
        \State \texttt{    ch[x:(x + bs),y:(y + bs)] }
        \EndFor
         \State \texttt{c1+=1}
      \EndFor
   \State $ \text{\textbf{return }ch, index\_key} $
\EndProcedure
\end{algorithmic}
\end{algorithm}
\section{Experimental Analysis and Discussion}
\label{sec:rad}
A thorough experimental and analytical analyses were carried out on the proposed SePriS system. A remote server and virtual machines were employed for the experiment. Specifically, functional test, security tests on the encryption schemes, and extended validation on the BC-based solution were carried out.

\subsection{Functional Test}
\begin{figure}[b]
    \vspace{-10pt}
    \centering
        \includegraphics[width=0.40\textwidth]{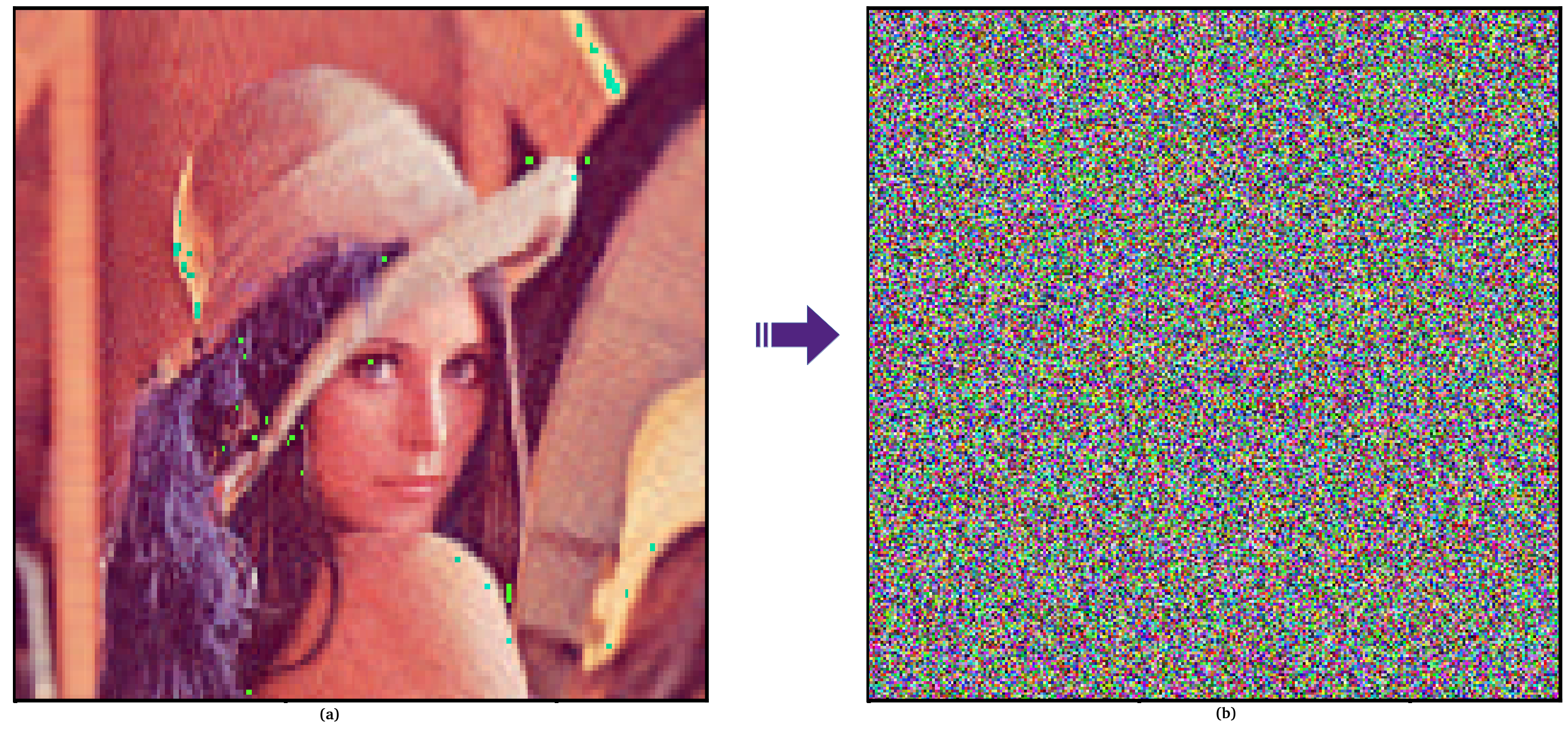}
    \caption{Enciphering: (a) Plain frame/Image, (b) AES Encrypted \& Shuffled}
    \label{enced}
\end{figure}

Figure \ref{enced} demonstrates the functionality of the proposed DAB scheme proving that it works as required. The plain image in Fig. \ref{enced}(a) is completely transformed into a random cipher in Fig. \ref{enced}(b) by the AES coupled with a BS Algorithm \ref{sfl}. 

\begin{figure*}[t]
    \centering
        \includegraphics[width=0.8\textwidth]{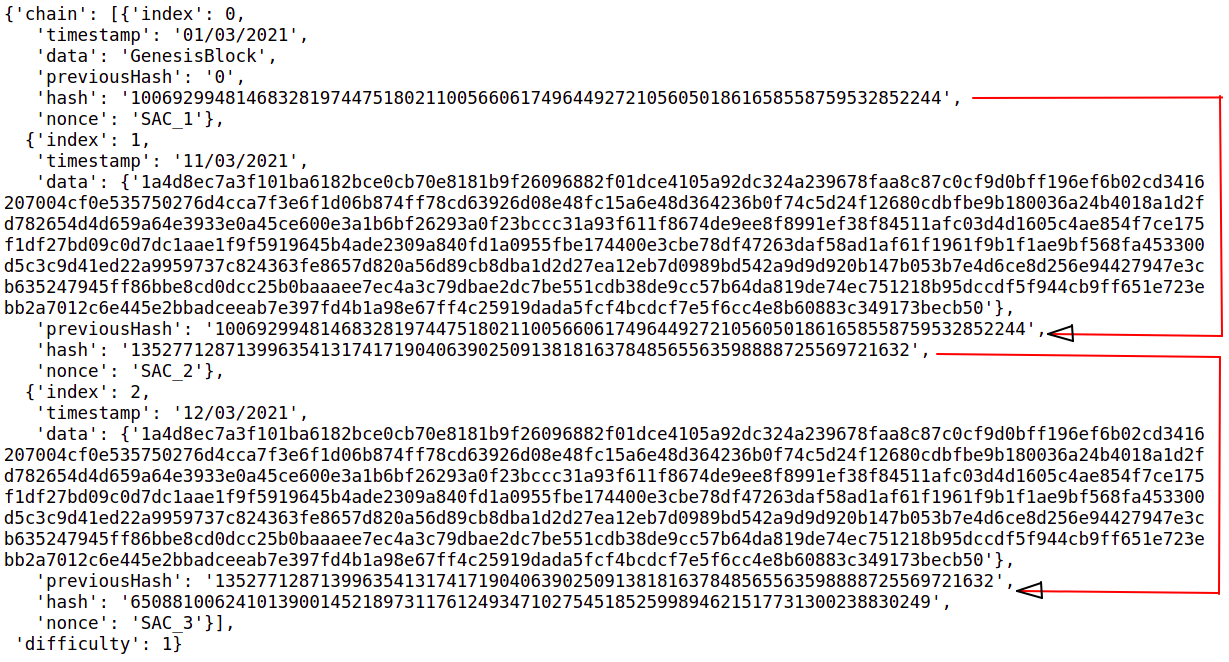}
    \caption{Creation of sample blocks with the data part enciphered: data comprises requests made and associated identify information and log of activities.}
    \label{blkc}
    \vspace{-10pt}
\end{figure*}

Figure \ref{blkc} shows sample blocks created when two requests for access to stored surveillance videos were made on two different dates. The request and all associated information are stored in the data part of the blocks. As shown by Fig. \ref{blkc}, the data section of the blocks are stored as cipher-text for security and privacy reasons. The secret keys that are used to unlock the data part are kept in the BC-nodes indexed and encrypted with their private keys.

\subsection{Security of the Enciphering Scheme}
To measure the security and computational performance of the proposed DAB scheme, a number of computational security and statistical parameters are considered. The tests or parameters considered include Encryption Quality, Frequency Test, Run Test, Gap Test, Poker Test, Pixel Sensitivity, Key Sensitivity, Peak Signal to Noise Ratio (PSNR), Entropy Test, and Correlation Analysis. 

Encryption Quality (EQ), as stated in Eq. \ref{eq:eq}, is a measure of changes in pixels after a plain frame/image has been encrypted. The higher the number of pixels changed, the better the quality of the enciphering scheme is. The proposed DAB scheme has an Encryption Quality of $99.998\%$, which means that almost every pixel in the plain image has a different value after encryption was performed. More details of Frequency Test, Run Test, Gap Test, and Poker Test are available to interested readers in the US National Institute of Technology (NIST) randomness test suite \cite{rukhin2001statistical, strube1983tests}. For detailed explanation of Pixel Sensitivity, Key Sensitivity, Peak Signal to Noise Ratio (PSNR), Entropy Test, and Correlation Analysis, please refer to \cite{fitwi2021privacy}. The results of all tests conducted are shown in Table \ref{tab:secenc}, which validate the the proposed DAB scheme.

\begin{equation}   
    \label{eq:eq}
      \begin{split}
        EQ\ &=\ \frac{Number\ of\ affected\ pixels}{Total\ Number\ of\ pixels}
      \end{split}
\end{equation}

\begin{table}[ht]
\centering
\caption{Security Analysis}
\label{tab:secenc}
\begin{center}
\begin{tabular}{|l|p{0.93cm}|p{3.5cm}|p{0.93cm}|}
\hline
\rule[-1ex]{0pt}{3.5ex} \textbf{Parameter/Test} & \textbf{Result} & \textbf{Remark}  \\
\hline\hline
\rule[-1ex]{0pt}{3.5ex} \textbf{Encryption Quality} & $99.998\%$ & $Ratio\ of\ changed\ pixels$ \\
\hline
\rule[-1ex]{0pt}{3.5ex} \textbf{Frequency Test} & $0.643$ & $P\_val\ >\ decision\ rule$ \\
\hline
\rule[-1ex]{0pt}{3.5ex} \textbf{Run Test} & $0.23$ & $P\_val\ >\ decision\ rule$ \\
\hline
\rule[-1ex]{0pt}{3.5ex} \textbf{Gap Test} & $Passed$ & $All\ p\_values\ >\ 0.05$ \\
\hline
\rule[-1ex]{0pt}{3.5ex} \textbf{Poker Test} & $Passed$ & $All\ p\_values\ >\ 0.05$ \\
\hline
\rule[-1ex]{0pt}{3.5ex} \textbf{Pixel Sensitivity} & $---$ & $To\ be\ secure:$ \\
\rule[-1ex]{0pt}{3.5ex} \textbf{ NPCR} & $99.151\%$ & $Must\ be\ >\ 99\%$ \\
\rule[-1ex]{0pt}{3.5ex} \textbf{ UACI} & $33.139\%$ & $Must\ be\ > 33\%$ \\
\hline
\rule[-1ex]{0pt}{3.5ex} \textbf{Key Sensitivity} & $---$ & $To\ be\ secure:$ \\
\rule[-1ex]{0pt}{3.5ex} \textbf{ NPCR} & $99.634\%$ & $Must\ be\ >\ 99\%$ \\
\rule[-1ex]{0pt}{3.5ex} \textbf{ UACI} & $33.548\%$ & $Must\ be\ > 33\%$ \\
\hline
\rule[-1ex]{0pt}{3.5ex} \textbf{PSNR Test} & $11.73dB$ & $Good\ if\ less\ than\ 20 dB$ \\
\hline
\rule[-1ex]{0pt}{3.5ex} \textbf{Entropy Test} & $7.945$ & $Good\ if\ closer\ to\ 8$ \\
\hline
\rule[-1ex]{0pt}{3.5ex} \textbf{Correlation} & $---$ & $Good\ if\ closer\ to\ 0$ \\
\rule[-1ex]{0pt}{3.5ex} \textbf{        Horizontal} & $0.0027$ & $---$ \\
\rule[-1ex]{0pt}{3.5ex} \textbf{        Vertical} & $0.0065$ & $---$ \\
\rule[-1ex]{0pt}{3.5ex} \textbf{        Diagonal} & $0.00781$ & $---$ \\
\hline  
\end{tabular}
\end{center}
\vspace{-10 pt}
\end{table}

\subsection{Security \& Privacy of SePriS system}
In order to ensure a secure and privacy-aware surveillance practice while sharing or accessing stored surveillance videos, the following requirements are set. 

\noindent\textbf{1. Security}: It is often viewed in terms of the confidentiality, integrity and availability (CIA) Triad. The confidentiality ensures that only authorized users can access the data under protection. The integrity property requires that data must be accurate in transit and be altered only by authorized parties. At last, the availability attribute of security states that the security mechanism in use must allow to access the data to the right users at the right time in the right way.

\noindent\textbf{2. Privacy}: It requires that the privacy of individuals caught on CCTV cameras and stored on off-BC sites must be preserved in optimal balance with usability.

\noindent\textbf{3. Authenticity}: the proposed scheme must have a capability to verify the identities of requestors before any access is granted.

\noindent\textbf{4. Accountability}: The scheme must have a capability for auditing users to hold them responsible for any misbehaving. 

\noindent\textbf{5. Auditability}: The proposed scheme must be auditable, which is an essential component of security. Logging and audit logs are, for instance, vital. The audit logs contain information on who have accessed which videos and what they did on them. 

\noindent\textbf{6. Anonymity}: This requirement refers to the fact that parties must have no visible identifier for privacy reasons. As our goal is to prevent abuse of surveillance videos by people in charge, our scheme does not guarantee complete anonymity to users who access stored video. However, individuals on video frames are anonymized! 

The proposed SePriS system is designed to meet the aforementioned requirements. As depicted in Table \ref{tab:bctest}, SePriS meets all requirements except anonymity, which is partially met. Within the secure group, in order to discourage leaking, some identity information about users are stored as part of the blocks in every BC-node in enciphered form.

\begin{table}[ht]
\centering
\caption{Security Analysis}
\label{tab:bctest}
\begin{center}
\begin{tabular}{|l|p{0.93cm}|p{3.5cm}|p{0.93cm}|}
\hline
\rule[-1ex]{0pt}{3.5ex} \textbf{Parameter} & \textbf{Result} & \textbf{Remark}  \\
\hline\hline
\rule[-1ex]{0pt}{3.5ex} \textbf{Security} & $Yes$ & $In\ terms\ of\ CIA\ Triad$ \\
\hline
\rule[-1ex]{0pt}{3.5ex} \textbf{Privacy} & $Yes$ & $Privacy\ is\ ensured!$ \\
\hline
\rule[-1ex]{0pt}{3.5ex} \textbf{Integrity} & $Yes$ & $By\ Double\ Lock\ Boxes$ \\
\hline
\rule[-1ex]{0pt}{3.5ex} \textbf{Authenticity} & $Yes$ & $Required\ before\ any\ access$ \\
\hline
\rule[-1ex]{0pt}{3.5ex} \textbf{Controllability} & $Yes$ & $Scheme\ is\ manageable$ \\
\hline
\rule[-1ex]{0pt}{3.5ex} \textbf{Auditability} & $Yes$ & $All\ activities\ are\ auditable!$ \\
\hline
\rule[-1ex]{0pt}{3.5ex} \textbf{Accountability} & $Yes$ & $Ensured\ by\ scheme!$ \\
\hline
\rule[-1ex]{0pt}{3.5ex} \textbf{Anonymity} & $No$ & $Partially$ \\
\hline
\end{tabular}
\end{center}
\vspace{-10 pt}
\end{table}
\section{Conclusions}
\label{sec:con}
Security and privacy are very essential issues in the world of video surveillance. Hence, this paper proposes SePriS, a private blockchain based solution coupled with an efficient video frame enciphering mechanism for sharing stored surveillance videos in a secure and privacy-aware way. It illustrates how the blockchain technology can be leveraged to prevent abuse and leaking of stored videos, which have plagued the surveillance practice for years. The experimental analysis show that the video frame enciphering mechanism passes the standard computational and security parameters. Additionally, they corroborate the security, privacy, integrity, authenticity, controllability, auditability, and accountability of the proposed decentralized system for sharing of stored surveillance videos. Overall, the SePriS system achieves the design goal and creates a video surveillance system with good balance of privacy and usability.
\ifCLASSOPTIONcaptionsoff
  \newpage
\fi
\bibliographystyle{IEEEtranS}
\bibliography{Reference.bib}
\end{document}